\documentstyle[preprint,aps,prb]{revtex}

\begin{document}
\draft
\title{Single-Electron Parametron: Reversible Computation in 
	a Discrete State System}
\author{Konstantin K. Likharev$^1$ and Alexander N. Korotkov$^{1,2}$}
\address{$^1$Department of Physics, State University of New York,\\
Stony Brook, NY 11794-3800 \\
and \\
$^2$Institute of Nuclear Physics, Moscow State University,\\
Moscow 119899 GSP, Russia}
\date{\today }
\maketitle

\begin{abstract}
We have analyzed energy dissipation in a digital device (``Single-Electron
Parametron'') in which discrete degrees of freedom are used for presenting
digital information. If the switching speed is not too high, the device may
operate reversibly (adiabatically), and the energy dissipation ${\cal E}$
per bit may be much less than the thermal energy $k_BT$. The energy-time
product ${\cal E}\tau $ is, however, much larger than Planck's constant 
$\hbar $, at least in the standard ``orthodox'' model of single-electron
tunneling, which was used in our calculations.
\end{abstract}



\narrowtext

\vspace{1ex}

Computation using any real physical system leads to the dissipation of
energy, because of unavoidable coupling between the degrees of freedom which
present information, and the environment. In most practical electronic
devices (such as semiconductor transistor circuits), energy is dissipated at
some rate even in static state, i.e.\ in the absence of information 
processing. Some prospective
digital devices, such as various Single-Flux-Quantum \cite{PQ76,RSFQ} and
Single-Electron \cite{Av-Likh,Kor} logics, \cite{QC} however, do not involve
static power consumption, because they present conservative systems where
digital information is coded by the choice of a local minimum of potential
energy. In this case the energy dissipation is proportional to the number of
logic operations. If such a conservative system is switched irreversibly
(e.g., as in the RSFQ logic \cite{RSFQ}), the energy loss ${\cal E}$ per one
logic operation is of the order of energy barrier $W$ separating the states.
The barrier should be sufficiently high to make the
probability $p\sim \exp (-W/k_BT)$ of thermally-induced errors low enough,
so that at the physically irreversible computation ${\cal E}_{\min }\sim
k_BT\ln (1/p)\gg k_BT$.

Some conservative systems, e.g.\ the Parametric Quantron, \cite{PQ76,Goto}
are flexible enough to allow independent adjustment of the potential tilt
and barrier height. As has been shown earlier, \cite{Landauer,Bennett} this
flexibility allows {\it physically reversible} (adiabatic) switching of the
system with ${\cal E}\ll k_BT$, if the information content of the system
remains intact. If, however, information is lost during the operation 
({\it informationally irreversible} computation), the minimum energy loss 
is at least ${\cal E}_{\min }=k_BT\ln 2$ per each lost bit. Thus only at 
{\it completely} (physically {\it and} informationally) {\it reversible}
computation, ${\cal E}$ may be made much less $k_BT$. In this case, ${\cal E}$
scales as $\tau ^{-1}$, where $\tau $ is the switching time, so that the
product ${\cal E}\tau $ is fixed. \cite{Landauer}

${\cal E}\tau $ has the dimensionality of Planck's constant, so the natural
question is whether quantum mechanics imposes any fundamental lower bound on
this product. A quantitative analysis of a particular reversible system (the
Parametric Quantron) has shown \cite{Likharev82} that ${\cal E}\tau$ may be
much less then $\hbar $. The analysis has, however, used the assumption that
the potential energy is a function of a {\it continuous} degree of
freedom (in that particular case the Josephson phase $\phi $). To our
knowledge, a similar analysis has never before been carried out for any
system presenting information with {\it discrete} states.

Recently we suggested \cite{param1} a system (``Single-Electron
Parametron'') based on correlated single-electron tunneling (see, e.g.\
Ref.\ \onlinecite{Av-Likh}), which may be used for reversible computation
using a discrete degree of freedom -- the electric charge $Q$. The goal of
the present work was to find the minimum energy dissipation ${\cal E}$ for
this system, and relate it to the switching time $\tau $ and to the error
probability $p$. We have shown, in particular, that within the ``orthodox''
theory of single-electron tunneling, \cite{Av-Likh91} ${\cal E}\tau $ is
always considerably larger than $\hbar $.

Figure 1a shows the possible structure of a unit cell of the system. It
consists of at least three small conducting islands (with capacitances $C\ll
e^2/k_BT$), with the middle island slightly shifted in $y$-direction. Tunnel
barriers with small conductances ($G\ll e^2/\hbar $) allow direct transfer
of electrons only between the neighboring islands. The system is biased by a
periodic ``clock'' electric field ${\bf E}(t)$ perpendicular to axis $x$.
Let us consider the conceptually simplest case when the cell is charged as a
whole by a single extra electron. (For practice, the operation using
electron-hole pairs may be beneficial, \cite{param1,param2} but for our
present discussion both versions are identical.) 

When the vertical component $E_y(t)$ of the field is lower than a certain
threshold value $E_t$, the extra electron is kept inside the middle island.
The energy diagram for this ``OFF'' state is shown at the top of Fig.\ 1b.
As $E_y$ is increased, tunneling of the electron from the middle island into
one of the edge islands becomes energy advantageous at $E_y>E_t$. If the
system is completely $x-$symmetric, this results in spontaneous symmetry
breaking, so that the direction of the resulting electric dipole moment of
the system is random: $P_x=Qd_{eff}, $ $d_{eff}\approx d$, $Q=\pm e$. However,
if the symmetry is broken by a weak additional external field $E_x$ (say,
induced by the dipole moment of a similar neighboring cell), the direction
of the electron tunneling and hence the sign of $P_x$ is predetermined by
this field. The middle frame in Fig.\ 1b shows this ``OFF$\rightarrow $ON
switching'' stage. Finally, when $E_y$ is well above $E_t$, the electron is
trapped inside one of the edge islands, even if the ``signal'' field $E_x$
now favors its transfer in the opposite edge direction (``ON'' state, the
bottom frame in Fig.\ 1b). In this state, electron transfer may only be
achieved via a higher-order ``co-tunneling'' process; the probability of
this process may be made negligibly small by either decreasing the tunnel
conductance or by inserting a few additional islands into the cell.
\cite{Av-Likh91} If this parasitic process is negligible, the cell has a 
fixed dipole moment and may serve as a robust source of signal field $E_x$ for
similar neighboring cells. After this source has been used, the system is
reset into the ``OFF'' state during the corresponding part of the clock
cycle, when $E(t)$ drops below $E_t$ again.

It is evident that the operation of the cell is quite similar to that of the
Parametric Quantron, \cite{PQ76} except now the information is presented by a
discrete variable, $Q=\pm e$. Similarly to the Parametric Quantron, the
Single-Electron Parametron may be used for reversible transfer and
processing of information. \cite{param1,param2} For example, Figure 2 shows a
possible structure of a shift register. In each neighboring cell, the extra
charge sign alternates, while the direction of the middle island shift
within plane $yz$ is changed by $\Theta =\pi -2\pi /M$, $M>2$ (in Fig.\ 2, 
$M=3$). The clock field ${\bf E}(t)$ has a fixed magnitude $E>E_t$, but
rotates within plane $yz$, providing periodic switching ON and OFF of the
cells, with the phase shift $2\pi /M$ between the neighboring cells. At an
appropriate choice of $E$ and the distance between the cells, \cite{param2}
the orientation of the dipole moment of the cells in ON state determines the
direction of the field $E_x$ and hence the direction of electron tunneling
in the neighboring cell which is being switched OFF$\rightarrow $ON. As a
result, the information is being re-written from cell to cell, and thus
transferred over $M$ cells each clock period. Reversible logic operations
may be implemented in a similar way, e.g.\ by using majority gates with
additional output cells. \cite{Likharev82}

Within the ``orthodox'' theory \cite{Av-Likh91} all properties of the system
may be found from solution of the system of master equations for the
probabilities $p_i(t)$ to find the extra electron in the middle $(i=m)$,
left $(i=l)$, and right $(i=r)$ islands: 
\begin{equation}
\label{masteq}\frac d{dt}p_i=\sum_{j=m,l,r}(p_j\Gamma _{ji}-p_i\Gamma
_{ij}),\,\,\sum_ip_i(t)=1,\,\,\,p_m(0)=1, 
\end{equation}
where in our case the tunneling rate matrix ${\bf \Gamma }$ has only four
nonvanishing components: 
\begin{equation}
\label{gamma}\Gamma _{mr}^{\pm }=\frac{\pm GW}{e^2\{1-\exp [\mp W/k_BT]\}}%
,\,\,\Gamma _{ml}^{\pm }=\frac{\pm G(W-\Delta )}{e^2\{1-\exp [\mp (W-\Delta
)/k_BT]\}}. 
\end{equation}

Here $\Gamma _{ij}^{+}\equiv \Gamma _{ij},\Gamma _{ij}^{-}\equiv \Gamma _{ji}
$, while $W(t)\approx E_y(t)d^{\prime }+const$ is the energy difference
between the charge configurations with the extra electron in the middle and
right islands. We will accept that near the decision-making point of the 
ON$\rightarrow$OFF switching ($t\approx 0$ in Fig.\ 1c) the difference is a
linear function of time: $W=\alpha t$. $\Delta \approx 2dE_x$ is the energy
difference between the left and right ON states. In order to operate with a
low error probability $p\ll 1$, at the decision-making moment this
difference should be large enough. Without the loss of generality, we may
assume $\Delta >0$; then $p$ can be found from the solution of the master
equation as $p_l(\infty )$, while the average energy ${\cal E}(t)$
dissipated by moment $t$ can be calculated as 
\begin{equation}
\label{W}{\cal E}(t)=\int_{-\infty }^t\{W(t)\frac{dp_r}{dt}\,+[W(t)-\Delta ]
	\frac{dp_l}{dt}\}dt\,;
\end{equation}
we will be mostly interested in the net dissipation ${\cal E}\equiv 
{\cal E}(\infty )$.

The solution of equations (\ref{masteq})--(\ref{W}) yields the following
results. With an accuracy sufficient for our final result, the total error
probability $p$ may be calculated as a maximum of probabilities of the 
{\it thermal} and {\it dynamic} errors. The thermal error may occur due to
thermally-activated tunneling to the wrong state (in our case, $l$), and its
probability $p_{therm}$ may always be expressed as $\exp (-\Delta /k_BT)$.
(Because of that, we will restrict our discussion to the limit $\Delta \gg
k_BT$.) The dynamic error occurs when the switching speed $\alpha $ is too
high, and the system remains in the initial (symmetric) state up to the
moment when tunneling to the upper energy level becomes possible. If $\delta
\equiv \alpha e^2/G\Delta k_BT\gg 1,$ the dynamic error dominates and its
probability is given by expression 
\begin{equation}
p_{dyn}=K\gamma \exp (-\frac 1{2\gamma }), \,\, 
K=\frac 1{2\gamma } - \frac{\sqrt{\pi }}{4\gamma ^{3/2}}
   \exp (\frac 1{4\gamma })[1-\mbox{Erf}(\frac 1{2\sqrt{\gamma }})]=
   1+\sum_{n=1}^\infty \frac{(2n+1)!}{n!}\,(-\gamma )^n,
\label{dyn} \end{equation}
where $\gamma \equiv \alpha e^2/G\Delta ^2$. In order to keep $p_{dyn}\ll 1$,
 $\gamma $ should be much smaller than 1, so that one can use Eq.\ 
(\ref{dyn}) with $K=1$.

Energy dissipation depends on another dimensionless parameter, $\beta \equiv
\alpha e^2/G(k_BT)^2=(\Delta /k_BT)^2\gamma =(\Delta /k_BT)\delta $. Like 
$\gamma $ and $\delta $, parameter $\beta $ is also proportional to the
switching speed $\alpha $, but is much larger than both of them (because 
$\Delta \gg $ $k_BT$) and may be comparable to unity. In the low-speed limit
$\beta \ll 1$, the switching process is adiabatic. It consists of numerous
tunneling events (back and forth between $m$ and $r$) taking place within
the energy interval $\sim k_BT$ around the point $W(t)=0$. In this case 
${\cal E}=\kappa \beta k_BT,$ where 
\begin{equation}
\label{kappa}\kappa =\int_{-\infty }^\infty \frac{e^x(e^x-1)}{x(1+e^x)^3}
	\,dx\,\simeq 0.426,
\end{equation}
so that for this (reversible) process ${\cal E}\ll k_BT$. Notice that 
${\cal E}=\kappa \beta k_BT=\kappa \alpha e^2/Gk_BT$ {\it decreases} when
temperature increases.

Our model allows not only to calculate the net dissipation ${\cal E,}$ but
also follow the time dynamics of energy transfer between the system and the
environment (``heat bath'') during the switching process. During the first
half of the process (when $W(t)\le 0$) the energy ${\cal E}_1\equiv -
{\cal E}(0)=T\ln 2\gg {\cal E}$ is borrowed from the heat bath (which, 
hence, is {\it cooled}), while virtually the same amount ${\cal E}_2 \equiv 
{\cal E}-{\cal E}(0)$ is returned back to the heat bath during the 
second half of the process ($W(t)\ge 0$). This exchange is directly 
related as ${\cal E}(t)=T\Delta S(t)$ to the temporal increase and 
consequent decrease of the entropy corresponding to the degree of freedom 
used to code information (in this particular case, the
polarization $P_x$). At the moment when $W=0$, the system may be in either
of two states ($p_m=p_r=1/2)$, i.e.\ $\Delta S=k_B\ln 2$ has been acquired
in the comparison with the definite initial state ($p_m=1$, $p_r=0)$. By the
end of the switching ($W\gg k_BT$) the entropy is restored to the initial
value since the state is definite again ($p_m=0$, $p_r=1$). Finite switching
speed decreases ${\cal E}_1$ and increases ${\cal E}_2$ (see the dotted 
lines in Fig.\ 3).

In the limit $\beta \gg 1$, the speed of energy change is so high that
switching may take place only at $W>0$, but within a much larger interval of
energies: $\Delta W\sim \beta ^{1/2}k_BT$. The average energy dissipation
for this (irreversible) process is of the same order, i.e.\ much larger than 
$k_BT$ and independent of temperature: ${\cal E}=(\pi \beta
/2)^{1/2}k_BT=(\pi e^2\alpha /2G)^{1/2}$. The results of numerical
calculation of ${\cal E}$ for intermediate values of $\beta $ are presented
by the solid line in Fig.\ 3.

Taking into account that at $p \ll 1$ the parameter $\tau =\Delta /\alpha $
may be considered as the duration of the switching process (Fig.\ 1c), all
our asymptotic results may be summarized as follows:

\begin{equation}
\label{et}{\cal E}\tau =\frac \hbar {GR_Q}\times \left\{ 
\begin{array}{l}
0.67\,\ln \frac 1p\,,\,\text{ for }\delta ,\beta \ll 1, \\ 
1.97\,\beta ^{-1/2}\ln \frac 1p\,,\,\text{ for }\delta \ll 1\ll \beta , \\ 
2.78\,(\ln \frac 1{2p\ln
(1/p)})^{1/2}\,,\,\text{ for }1\ll \delta ,\beta \, ,
\end{array}
\right. 
\end{equation}
where $R_Q$=$\pi \hbar /2e^2\approx 6.45 \, \mbox{k}\Omega$ is the 
quantum unit of
resistance. Since the orthodox theory is valid only at $GR_Q\ll 1$, within
this theory ${\cal E}\tau \gg \hbar $ for any switching speed.

To summarize, we have shown that reversible computation with the energy
dissipation ${\cal E}$ per bit well below $k_BT$ may be implemented in a
physical system with discrete states. The quantum bound for the product
${\cal E}\tau $, obtained within our concrete model is, however, much higher
than that obtained earlier for a system with continuous degrees of freedom.
\cite{Likharev82} Apparently the $\hbar $-limit for ${\cal E}\tau $ may be
overcome in the case of islands with discrete spectra of electron energies, 
\cite{AvKor90,Kastner} though this may require an exponentially high energy
barrier $W$ during the ON state of the cell. A quantitative analysis of this
opportunity is in progress.

\vspace{1cm}  Useful discussions with D.\ V.\ Averin and T.\ Usuki are
	gratefully acknowledged. 
	The work was supported in part by ONR Grant \#N00014-93-1-0880
and AFOSR Grant \#F49620-95-1-0044.

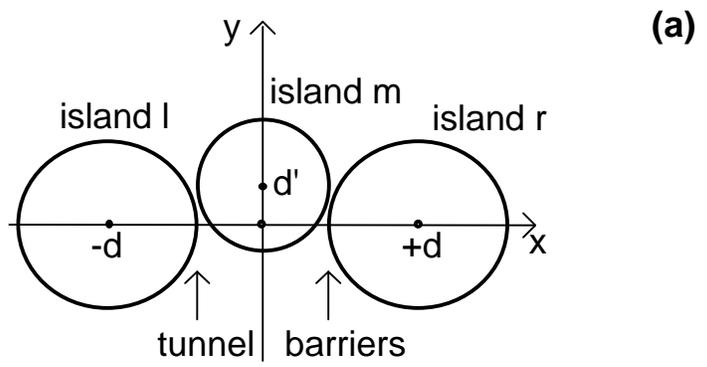
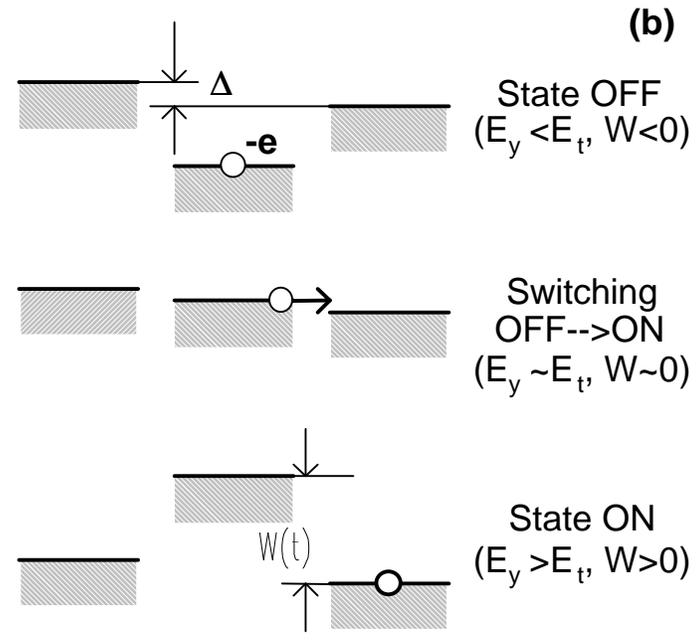
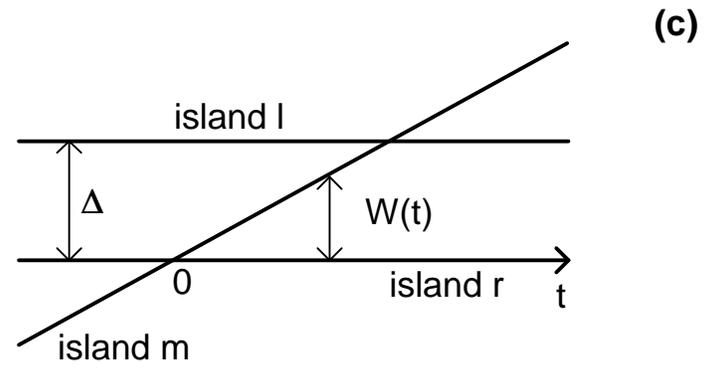
\begin{figure}
\caption{The Single-Electron Parametron: (a) 3-island version of 
the system, (b) its energy diagram for three  values of the clock field
$E_y$, and (c) energy of the extra electron in various islands as a function
of time, close to the decision-making moment $t\approx 0$.  }
\label{Fig1}\end{figure} 

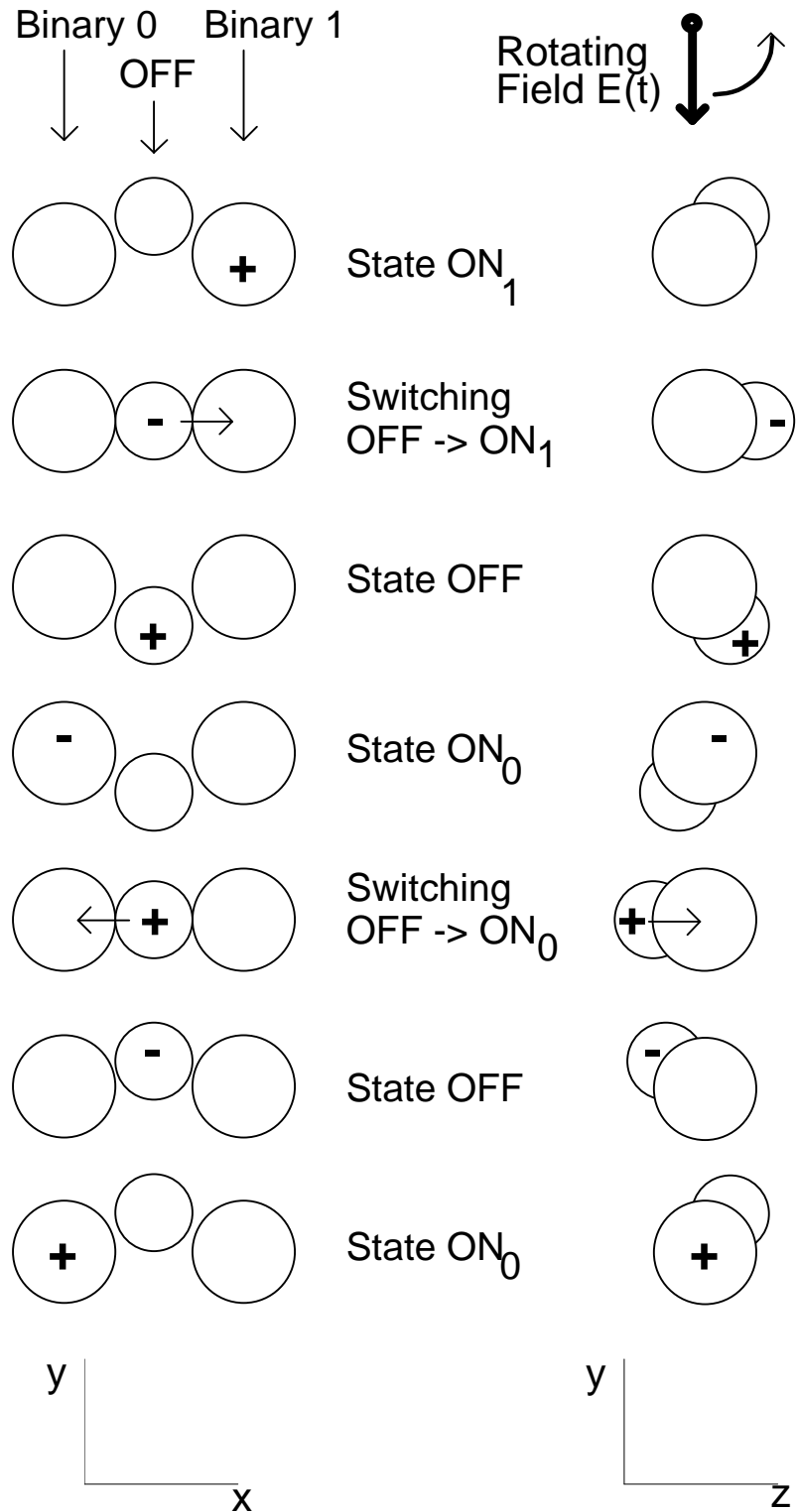
\begin{figure} 
\caption{ Top (left) and side (right) views of a shift register using an array
Single-Electron Parametron cells. Clock field $E(t)$ rotates in $yz$ plane. 
Digital bits are coded by positions of the extra charges in ON state 
of the cells,
and are propagated from the top to the bottom, over $M=3$ cells during 
one clock period. } 
\label{Fig2}\end{figure}  

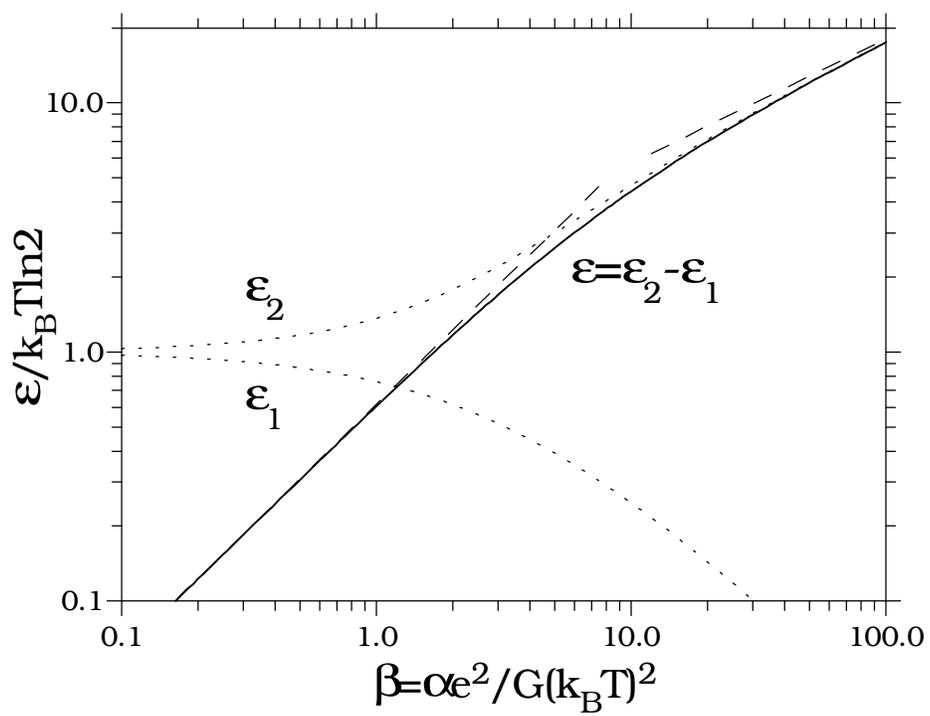
\begin{figure} 
\caption{ Components of the energy exchange between the Parametron 
and the heat
bath as functions of the process speed $\alpha=dW/dt$. Dotted lines: average 
energy flow ${\cal E}_1$ from
the heat bath to the Parametron during the first half of the process 
($W\le 0$) and the average flow
${\cal E}_2$ from the device back into the heat bath during its second 
half ($W\ge 0$), respectively.  Solid line: net energy dissipation ${\cal E}
={\cal E}_2-{\cal E}_1$. Dashed lines show the low-speed (adiabatic) 
and high-speed (diabatic) asymptotes of the function ${\cal E}(\alpha)$ -- 
see formulas in the text. }
\label{Fig3}\end{figure}  

\end{document}